\documentclass[12pt, preprint]{aastex631}

\usepackage{savesym}
\savesymbol{tablenum}
\usepackage{siunitx}
\restoresymbol{SIX}{tablenum}
\usepackage{longtable}
\usepackage{graphicx,epstopdf}
\usepackage{comment}

\DeclareGraphicsExtensions{.ps}
\usepackage{amsmath,siunitx}
\DeclareUnicodeCharacter{2212}{-}

\newcommand{\pan}{Pan-STARRS}

\newcommand{\msun}{\hbox{${M}_{\odot}$}}

\newcommand{\h}{$h_{70}^{-1}$}
\newcommand{\pant}{Pan-STARRS $3\pi$ survey}
\newcommand{\be}{\begin{equation}}
\newcommand{\ee}{\end{equation}}
\newcommand{\ba}{\begin{eqnarray}}
\newcommand{\ea}{\end{eqnarray}}

\newcommand{\swift}{\emph{Swift}}

\newcommand{\simgt}{\lower 2pt \hbox{$\, \buildrel {\scriptstyle >}\over {\scriptstyle\sim}\,$}}
\newcommand{\simlt}{\lower 2pt \hbox{$\, \buildrel {\scriptstyle <}\over {\scriptstyle\sim}\,$}}
\newcommand{\ls}{\lower 2pt \hbox{$\;\scriptscriptstyle \buildrel<\over\sim\;$}}
\newcommand{\gs}{\lower 2pt \hbox{$\;\scriptscriptstyle \buildrel>\over\sim\;$}}

\graphicspath{{./}{figures/}}

\received{May 30, 2021}
\accepted{November 17, 2021}

\submitjournal{ApJS}

\shortauthors{Bhatiani et al.}
\graphicspath{{./}{figures/}}
\begin{document}

\title{\large Optical Confirmation of X-ray selected Galaxy clusters from the Swift AGN and Cluster survey with MDM and Pan-STARRS Data (Paper III)}

\correspondingauthor{Saloni Bhatiani}
\email{salonibhatiani@ou.edu}

\author[0000-0002-9044-9383]{Saloni Bhatiani}
\affil{Homer L. Dodge Department of Physics and Astronomy,
University of Oklahoma, Norman, OK 73019, USA}

\author[0000-0001-9203-2808]{Xinyu Dai}
\affil{Homer L. Dodge Department of Physics and Astronomy,
University of Oklahoma, Norman, OK 73019, USA}
\email{xdai@ou.edu}

\author{Rhiannon D. Griffin}
\affil{Homer L. Dodge Department of Physics and Astronomy,
University of Oklahoma, Norman, OK 73019, USA}

\author{Jenna M. Nugent}
\affil{Homer L. Dodge Department of Physics and Astronomy,
University of Oklahoma, Norman, OK 73019, USA}

\author[0000-0001-6017-2961]{Christopher~S.~Kochanek}
\affiliation{Department of Astronomy, The Ohio State University, 140 West 18th Avenue, Columbus, OH 43210, USA}
\affiliation{Center for Cosmology and AstroParticle Physics, The Ohio State University, 191 W.\ Woodruff Ave., Columbus, OH 43210, USA}

\author{Joel N. Bregman}
\affil{Department of Astronomy, University of Michigan, Ann Arbor, MI 48109, USA}



\begin{abstract}
To understand structure formation in the universe and impose stronger constraints on the cluster mass function and cosmological models, it is important to have large galaxy cluster catalogs. The Swift AGN and Cluster Survey is a serendipitous X-ray survey aimed at building a large statistically selected X-ray cluster catalog with 442 cluster candidates in its first release.  Our initial SDSS follow-up study confirmed $50\%$ of clusters in the SDSS footprint as z $<$ 0.5 clusters.  Here, we present further optical follow-up analysis of 248 (out of 442) cluster candidates from the Swift cluster catalog using multi-band imaging from the MDM $2.4m$ telescope and the Pan-STARRS survey. We report the optical confirmation of 55 clusters with $> 3\sigma$ galaxy overdensities and detectable red sequences in the color-magnitude space. The majority of these confirmed clusters have redshifts z $<$ 0.6. The remaining candidates are potentially higher redshift clusters that are excellent targets for infrared observations. We report the X-ray luminosity and the optical richness for these confirmed  clusters. We also discuss the distinction between X-ray and optical observables for the detected and non-detected cluster candidates.
\end{abstract}


\keywords{catalogs -- galaxies: clusters: general -- galaxies: groups: general -- large-scale structure of universe -- surveys}


\section{Introduction} \label{sec:intro}

Observational studies of the distribution of galaxies in the universe reveals inhomogeneity and structure on megaparsec and larger scales. Galaxy clusters and groups contain virialized assemblys of galaxies and they are the largest gravitationally bound structures with typical masses ranging from $10^{14}$ -- $10^{15}$~\msun. Studying them is significant for understanding the constitution and assembly history of these systems and probing the large-scale structure of the Universe \citep[e.g.,][]{bah83, bah88, bah97, carl96, pos86,pos92, ein97, bor01, zeh05}. %(Bahcall 1988; Carlberg et al. 1996; Bahcall et al. 1997) 
Statistical studies of galaxy clusters impose strong constraints on the cosmological parameters and cosmological models of the growth of structure \citep{voit05,allen11}. For example, weak gravitational lensing and X-ray observations provide constraints on cluster masses \citep{blain99,met03,smith05, okabe10, okabe11,okabe16, applegate14, hoe15}. %(Blain et al. 1999; Metcalfe et al. 2003;Smith et al. 2005;Okabe, Okura \& Futamase 2010; Okabe et al. 2011;Applegate et al. 2014;Hoekstra et al. 2015;Okabe et al. 2016)
The cluster mass function can then be used to constrain the dark energy equation of state \citep{munshi03,mantz14} %(Munshi et al. 2003; Mantz et al. 2014) 
 and neutrino masses \citep{carb12}. %(Carbone et al. 2012) 
  Galaxy clusters also provide a high density environment for studying galaxy formation, evolution and dynamics \citep{butch78, dre80, dress92,gar99, pog99, got03,smi05, pos05, von07,maug12, lau14}. %(Dressler \& Gunn 1992;Butcher \& Oemler 1978; Garilli et al. 1999; Goto et al. 2003; Maughan et al. 2012)
 
 Galaxy clusters can be observed across the electromagnetic spectrum and through gravitational lensing. These emissions correspond to different physical components of the cluster and lead to a variety of cluster detection techniques.  
 The detection of galaxy clusters using optical images was the first method used to build cluster catalogs %(Abell 1958; Zwicky, Corwin and Olowin 1989),
 and developed a statistical understanding of the cluster population \citep{abell58, zwicky61, abell89}. 
 The emergence of wide-field multi-band imaging surveys has led to the development of many cluster finding algorithms including galaxy density mapping \citep{maz07,adam10},
 %(Mazure et al. 2007; Adami et al. 2010) 
friends-of-friends algorithms \citep{huc82, li08,feng16},
%(Li et al. 2008; Feng et al. 2016
 and  Voronoi Tesselation methods \citep{ew93,ramella01,lopes04}.
%(Ebeling \& Weidenmann 1993, Ramella et al. 2001,Lopes et al. 2004; van Breukelen et al. 2009; Murphy et al. 2015).
One common optical detection method uses the tight color-magnitude relation of the early-type galaxies in the clusters to identify clusters \citep{gladders00,gladders05,castellon14}. %(Gladders \& Yee 2000, Gladders \& Yee 2005; Nilo Castellon et al. 2014)
 %The quiescent stellar population in these galaxies develop strong metal absorption lines blueward of 4000 Angstrom which appears as a break in their spectra. In the color-magnitude space, this break manifest as a distinct feature called the red sequence. 
 Several cluster finders based on the cluster red sequence method have yielded large cluster catalogs within the SDSS and Dark Energy Survey, such as maxBCG \citep{koe07}, GMBCG \citep{hao10}, AMF \citep{szabo11,ban18}, WHL2012 \citep{wen12}, and redMaPPer \citep{ryk14,ryk16}. 
However, optical cluster finding algorithms suffer from projection effects as galaxy clusters are three dimensional objects that are projected on a 2D sky, especially at higher redshifts as the contamination from the foreground galaxies increases. %\textbf{Although larger efforts have been focused on cluster surveys in the optical-NIR and X-rays, in the last decade, cluster detection by Sunyaev-Zel’dovich (SZ) signatures in the microwave has emerged as a promising detection technique that is sensitive to high mass, high redshift galaxy clusters \citep{mcin09,brod10,hinc10,vand10,fole11,plan11b, mena12,stal12, hass13}. The thermal SZ effect is caused by the inverse Compton scattering of the Cosmic microwave background (CMB) photons by the high energy electrons contained within the intracluster medium \citep{suny72,suny80,carl02}. This effect manifests as distortions in the CMB spectrum that are untouched by cosmic dimming, and the magnitude of the effect shows a correlation with the cluster mass \citep{bona08,marr12,sifo12,plan13}, thus enabling a redshift-independent detection of all clusters above the mass threshold set by the survey. Nonetheless, because of the redshift-independent nature of the SZ signals, follow-ups in the optical are necessary for verification and determination of redshifts.}

The intracluster medium (ICM) of galaxy clusters is hot plasma that produces X-ray emission by the thermal bremsstrahlung process \citep{felt66,mit76,bs77,kb12}. 
As bright extended sources, clusters are easily identified in X-ray surveys and they stand out from the background because the emission is proportional to the square of the electron number density \citep{voit05,eb98}. X-ray selection also characterises the hot intracluster gas component that accounts for the majority of the baryonic mass of the cluster \citep{cavff76,all02} yielding cluster samples with well-characterized cluster masses. 
Studies have suggested that the X-ray luminosity and mass correlation is tighter than that between optical richness and mass relation so that X-ray methods provide more accurate measurements of cluster masses  \citep{bohr20,voit05}.
A slew of X-ray cluster surveys with varying energy range, depth, and area have been conducted including the Northern ROSAT All-Sky Survey \citep[NORAS,][]{bohr20}, the ROSAT-ESO Flux Limited X-ray Cluster Survey \citep[REFLEX,][]{bohr01}, the Massive Cluster Survey \citep[MACS,][]{eb01}, and the Highest X-ray flux Galaxy Cluster Sample \citep[HIFLUGCS,][]{reip02}. More recent surveys are based on XMM-Newton and Chandra observations and include the XMM-Large Scale Structure survey \citep[XLSS,][]{pac07}, the Chandra Multiwavelength Project Serendipitous Galaxy cluster survey \citep[ChaMP,][]{kim04,green04,bark06}, and the 3XMM/SDSS Stripe 82 galaxy cluster survey \citep{takey16}. These X-ray surveys have uncovered a sizable sample of galaxy clusters extending up to photometric redshifts of 1.9 \citep{basil04,pop04,piff11,mehr12,clerc12,takey11,takey13,takey14}. With the advent of the next generation of all-sky X-ray survey, eRosita, we can expect to detect several hundred thousand clusters \citep{pill12}. The hot X-ray emitting gas also introduces Sunyaev-Zel’dovich \citep[S-Z;][]{suny72,suny80,carl02} distortions in the microwave background that can be used to identify clusters  \citep{mcin09,brod10,hinc10,vand10,fole11,plan11b, mena12,stal12, hass13}. Since the S-Z effect is a scattering effect that is based on absorption of energy, it has the advantage that the signal amplitude is nearly independent of distance although the optical survey resolution does depend on redshift.

 The Swift AGN and cluster survey (SACS) is a serendipitous soft X-ray survey \citep{dai15}. It is a wide-field survey spanning an area of 125 square degrees in the sky with a median flux limit of $10^{-15}$ erg cm$^{-2}$ s$^{-1}$. SACS targets Gamma-ray burst (GRB) fields that are randomly distributed across the sky and have no correlation with known X-ray sources. Thus, SACS is a medium deep, broad-field, serendipitous X-ray survey that is ideally suited for detecting galaxy clusters and AGNs at intermediate redshifts. The first release of the survey yielded a total of 442 cluster candidates \citep{dai15}, which require a multiwavelength investigation to establish their properties. Despite the many advantages of X-ray surveys over optical surveys, the X-ray detection method pose some limitations.  While X-ray probes favor massive systems with deep potential wells,  the low mass and gas-poor clusters remain hidden and surface brightness dimming makes it difficult to detect high redshift clusters. The biggest limitation, however, is that optical observations are almost always required to determine the redshifts. Approximately 25 square degrees of the SACS area overlapped with the SDSS DR8 survey \citep{aih11} so the initial optical follow-up was conducted using the SDSS archival data \citep{griff16}. Out of the 442 SACS cluster candidates, 209 fell in the footprint of SDSS DR8 and 103 were confirmed as galaxy overdensities with a red sequence methods that yielded a photometric redshift in the redshift range of $z<0.8$, where the cluster sample is complete below $z<0.3$ and 40\% and 25\% complete at $z=0.5$ and $z=0.8$. 
 The redshift distribution of the SDSS confirmed clusters is consistent with the theoretical predictions for SACS given its X-ray flux limits and models for the cluster mass function \citep{tink08}. \citet{griff16} found that about 30\% of the cluster candidates that fell in the SDSS regions were low redshift clusters ($z<0.5$), 14\% were recognized as $0.5<z<0.8$ clusters, and 
 the remaining unconfirmed candidates likely have redshifts $ z\gtrsim 0.3$ \citep{griff16}.
 %On comparing the lower redshift distribution with the model, the survey was found to be 80\% complete for z$\lesssim$0.4. This means that the SDSS covers a majority of the clusters with redshifts up to 0.5, however, on account of the shallowness of the SDSS observations, the distribution of the SDSS confirmed clusters is observed to be considerably lower at higher redshifts. 

We have now performed optical follow-up observations with the MDM 2.4m Hiltner, KPNO 4m Mayall, and CTIO 4m Blanco telescopes, and used public \pan \ and DES survey data to further study the \swift\ cluster candidates. These observations are both deeper images of the SDSS regions and expansions to cover the non-SDSS regions.
In this paper, we present results from MDM/Hiltner and \pan \ (north of $-30^{\circ}$ targets). The CTIO/Blanco and DES results will be presented in a companion paper. The layout of the paper is as follows. Section \ref{sec:follow} describes the optical follow-up data for this paper. In section \ref{sec:analysis}, we discuss how we verify the \swift\ cluster candidates using the optical over-density/red sequence method.  In section \ref{sec:discuss}, we end with the conclusions and a discussion of our results.  
We assume cosmological parameters of $\Omega_M = 0.27$, $\Omega_{\Lambda}=0.73$, and $H_0 = 70$~km~s$^{-1}$~Mpc$^{-1}$ throughout the paper.

%\begin{figure}
%    \includegraphics[width=0.5\textwidth]{limmag_pan.pdf}
%   \includegraphics[width=0.5\textwidth]{limmag_mdm.pdf}
%    \caption{Average limiting magnitude in $g$, $r$, $i$ bands for \swift\ cluster candidates in the \pant (left) and our observations with MDM (right). The dashed line represents the 5$\sigma$ limit.  
%    \label{maglim}}
%\end{figure}

\section{Optical follow-up Data}\label{sec:follow}

We primarily used MDM/Hiltner and Pan-STARRS data as the data obtained by the KPNO/Mayall was affected by sub-optimal observing conditions. While we observed 66 northern the \swift\ cluster candidates in 39 fields using the 4m Mayall, the observing conditions were non-photometric/partially cloudy, and we were unable to attain the expected photometric depths. Compared with the corresponding sources in \pant \ catalog, the magnitude limits of our Mayall images are 1--2 mag brighter.  Therefore, we used the $3\pi$ catalog in the subsequent analysis. \pan \ 1 (PS1) encompasses several surveys, two of which are of relevance here: the \pan \ $3\pi$ Steridian survey (DR1, \citet{cham16}; %(Chambers et al. 2016)  
DR2, \citet{flew18})
%(Flewelling 2018) 
, covers 30,000 square degrees of the sky north of $-30$ declination, and the Medium deep survey consisting of nightly observations of ten smaller fields distributed across the sky. Although, the $3\pi$ survey is a relatively shallower survey with $5\sigma$ depths of 23.3, 23.2, 23.1, 22.3, 21.3 in $g$, $r$, $i$, $z$, $y$, respectively, it's wide area means it includes most of the Swift clusters. We downloaded DR1 and DR2 source catalogs for 11 arcmins regions around the \swift \ cluster centers. 
%Figure~\ref{maglim} (left) shows the average limiting magnitude for the  procured Pan-STARRS fields. 

%These photometric depths are consistent with the expected depths for the detection of cluster candidates and are definitely better than the Kitt peak depths, therefore we use the 3 pi data for all targets that fall outside the purview of the SDSS. 

We also observed 53 \swift\ cluster candidates with the 2.4m Hiltner Telescope at the MDM observatory with OSMOS and either the blue or red 4K detector between 2011 to 2013.  These cluster candidates are all in SDSS, but unconfirmed in the SDSS archival analysis of \citet{griff16}. The images have a field of view of $11.5^2$ arcminutes on each side, and the seeing range between 1--2.5 arcsecs. For each target, we observe all the fields in the $g$, $r$, $i$ filters and a fraction in $z$ with 3--4 dithered images per filter.
%The R4K detector is fully depleted LBNL 4kx4k CCD with a thickness of 25 $\mu$m and 15 micron pixels. The plate scale of the detector is 0.17 arcseconds per pixel yielding a field of view of about 11.5 arcminutes on each side. 
%Deep Imaging of 70 Target fields was performed in the SDSS g, r, i bands with exposure times of .. secs respectively, to attain a limiting magnitude of ... with a S/N  of 10 for point sources. 
%These observations were interspersed with standard star field observations in order to obtain more accurate Magnitude Calibrations for the Science images.
Calibration data, including bias, sky or dome flats, were also obtained for each night of observation. 
We first performed overscan, bias, cross-talk, and flat-field corrections, then created super-flat images for fringing in the longer wavelength $i$ and $z$ images and updated the astrometry of the images using the USNO B1.0 catalog.
We used SWarp tool \citep{bertin02} to median combine the dithered images in each band, and generated a panchromatic image by combining all the images for each field.
Source detection and flux measurement was performed with SExtractor  \citep{ba96} in the dual image mode using the panchromatic image for detection and the band specific image for the fluxes.
Since all these MDM fields are in SDSS, the photometry calibration is performed relative to SDSS magnitudes.

\begin{figure}
 \includegraphics[width=1\textwidth]{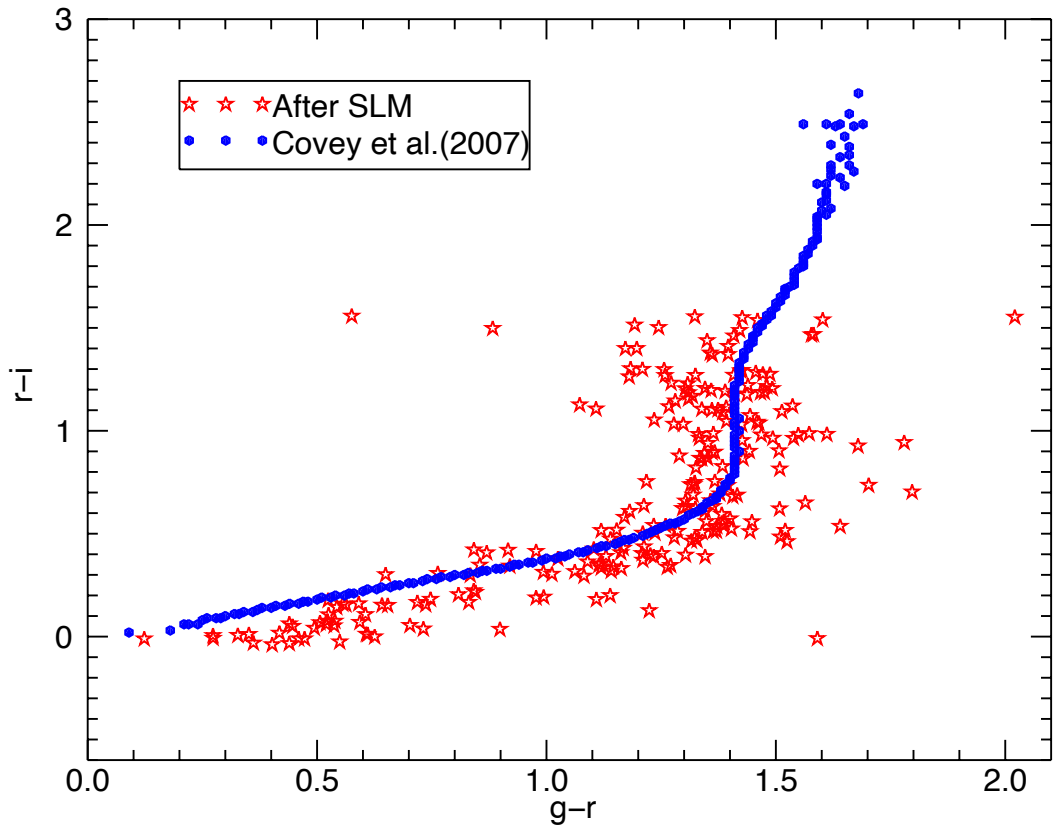}
    \caption{An illustration of the stellar locus matching technique employed to determine photometric color corrections for the MDM data. The standard stellar locus from \citet{cov07} is represented by the blue curve. The data points in red are the photometrically calibrated colors for stars in an MDM field. 
    \label{image1}}
\end{figure}

\newpage
\section{Optical Cluster Overdensity Analysis}\label{sec:analysis}

\subsection{Stellar locus correction}

Accurate measurements of photometric redshifts require robust photometric calibrations to accurately determine the photometric colors. We calibrate the colors using stellar locus regression \citep[SLR,][]{high09,ive07,des12, kelly14} for the MDM data. This technique is based on the known colors of the stellar main sequence. 
The Pan-STARRS colors have already been corrected based on this method \citep{high09}, and we have verified this from our independent analysis as well.
%We have utilized this universality of the stellar locus to calibrate the instrumental colors of the coadded images by matching the observed locus of the stars in the fields with the standard locus. 
We measure the stellar locus using the standard high-quality superclean sample of $\sim$500,000 stellar sources from \citet{cov07} jointly observed by the SDSS and Two Micron All Sky Survey (2MASS) surveys. The standard stellar locus exhibits a prominent kink feature at $g-r \sim 1.4$ and $r-i \sim 0.7$ in the ($g-r$, $r-i$) color plane, which we use as the main feature to perform calibration. 
The red side of the stellar locus is dominated by M dwarfs \citep{fin00,cov07,jur08,high09}, which are intrinsically dimmer compared to the more luminous stars on the blue $r-i < 0.7$ side of the stellar locus. 
Hence, to measure the entire locus including the kink, it was imperative to maximize the number of stars on each branch of the stellar locus. 
The stars used to identify the stellar locus were selected based on SExtractor's star/galaxy classifier parameter and a magnitude uncertainty of less than 2 mag to include enough faint stars on the red side of the kink for this analysis.
Although the individual measurement uncertainty is large, the mean trend can be constrained much better with the large sample of stars. 
We bin the stars by their $r-i$ color and the median of the $g-r$ color for each bin. Next, we perform a sigma clipping followed by a median smoothing of the two colors such that each point in the color-color space is replaced by the median in the closest windows of points. A typical field locus spans a color range of approximately 2 mags and the  typical color bin width considered is 0.02 magnitudes. We then fit a polynomial to the sequence of colors, identify the kink, and shift the colors to align with the calibration sequence. 

%The rms orthogonal scatter around the stellar locus is found to be --- in the (g-r, r-i) plane, incomparsion to the SDSS rm scattter of ---. 

%Our calibrated colors for MDM observations did not match the standard stellar locus, therefore, stellar locus matching was performed to determine the color corrections. For the Panstarrs data, the photometric colors matched the standard stellar locus reasonably and no further improvements were made.

\subsection{Redshift estimation using colors}
To find clusters we search for galaxy overdensities in three-dimensional space using both galaxy positions and the photometric colors or redshifts. We use a method that exploits the fact that the cluster galaxy population and the background have a bimodal color distribution \citep{hao10}.
%We first study these cluster candidates in the color-magnitude space to search for the red sequence feature, a ubiquitous signature of virialized clusters. 
We select galaxies by imposing cuts on the star/galaxy classifier and magnitude uncertainties. For the MDM fields, we required SExtractor parameters of $CLASS\_{STAR} < 0.8$ and $MAG\_AUTO_{err} < 0.33$ ($ S/N = 3$), while for \pan \ fields we required $ ipsf-ikron > 0.05$ and $ikron_{err} <0.3$. For each cluster candidate, galaxies were chosen within a source region of typical cluster size ranging from 1--2 Mpc. For the photometric depths of our data, we are primarily sensitive to $z \lesssim 0.4$, so the clusters over 2--3 arcmins in size. Hence we choose a  source radius of $\ang{;3;}$, and the background annulus from $\ang{;7;}$ to $\ang{;10;}$, both centered on the X-ray centroid position. 
The cluster candidates in the MDM data set are expected to be higher redshift clusters, so we used a source radius of $\ang{;1.5;}$, and a background annulus from $\ang{;5;}$ to $\ang{;10;}$, excluding regions within $\ang{;5;}$ from other cluster candidates in the field.
%For the Panstarrs, each field is 10 square arcmin in size and centered on the cluster candidate, therefore we have selected the source region as a circle of radius 2-3 arcmins 
%However, each MDM field is 10.5x10.5 arcmin2 and subsumes multiple cluster candidates, for that reason we have used an alternative approach to determine the background region for MDM fields. The source region for each MDM field is considered as a 1.5 arcmin region around the X-ray centroid of the cluster and an annular region extended up to 5 arcmins from the cluster center is considered as the no-zone region.
%The background counts were extracted for each Swift cluster per field from a circular region of radius 10 arcmins around the field center excluding the source and the no-zone regions corresponding to the Swift cluster.
%On studying these galaxies within the source search radius on the color-magnitude diagram, we were able to observe clear sequences in about one-fourth of the cluster candidates, nonetheless, this is a qualitative identification of galaxy clusters and does provide statistical significance of detection. Hence, 
We examine the color-distribution of these cluster candidates to identify galaxy overdensities and determine the significance of any detection. 
%Therefore, we compare the color distribution of the source region with the background region to determine galaxy overdensity on a color scale. 
 \begin{figure}
    \includegraphics[width=\textwidth]{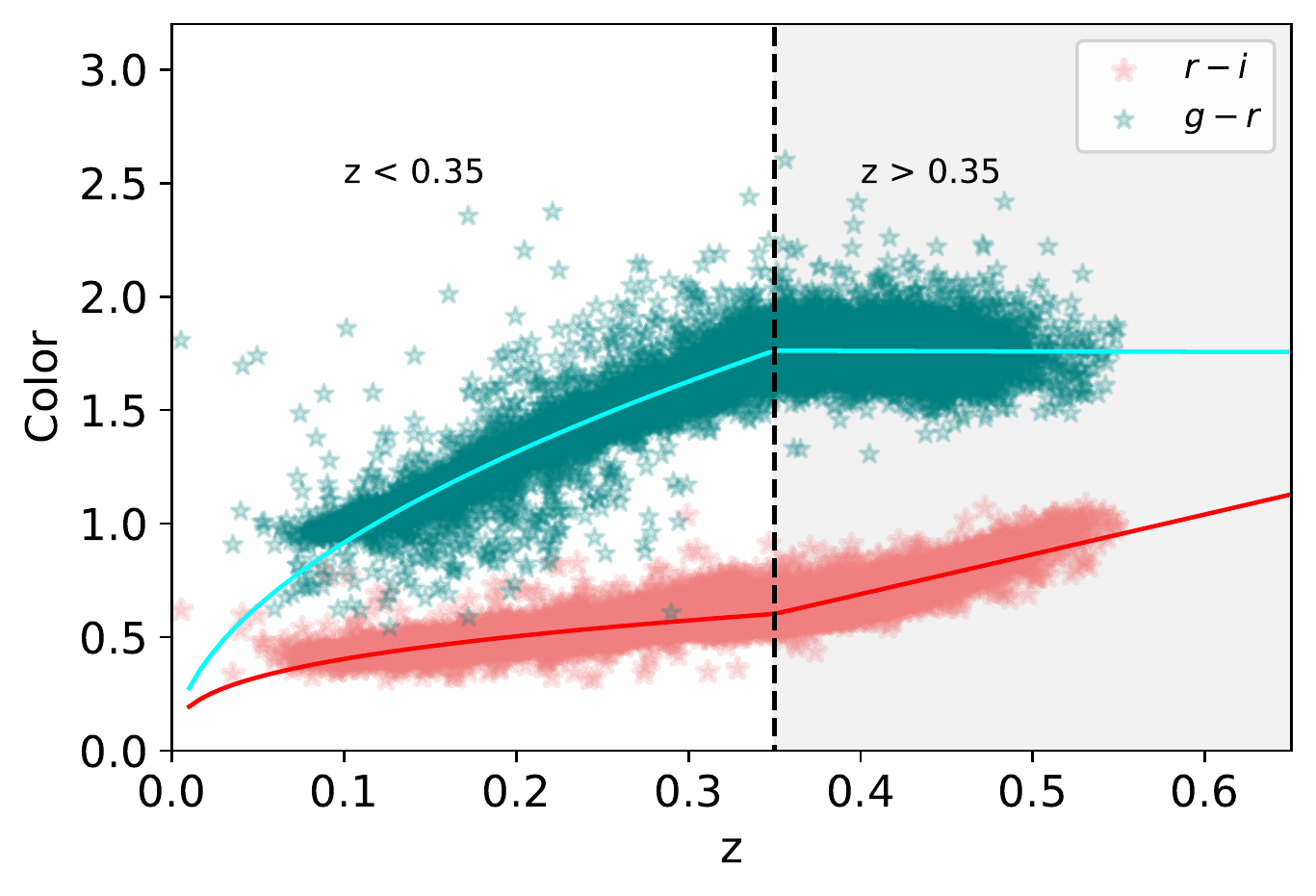}
\caption{Red sequence color as a function of the spectroscopic redshift for the galaxy clusters in the SDSS from the GMBCG catalog \citep{hao10}. A broken power law is fitted to $g-r$ (in blue) and $r-i$ (in red) with a break at $z=0.35$ (dashed black line).
 \label{image3}}
\end{figure}
 In order to perform a comparative study of the galaxy color distribution of the cluster and background, the background counts per bin were normalized to the source region using the ratio of the source to the background area. The background is used to estimate the contamination from interloping galaxies within the cluster region. To determine the over-density per bin, we compare the source count per color bin with the corresponding normalized background count. Assuming a Poisson distribution, we estimate the standard deviation of the over-dense bins as :
\begin{equation}
\sigma= \sqrt{N_{src} + N_{bkg}*{ \Bigg(\frac{A_{src}}{A_{bkg}}\Bigg)^2}}
\end{equation}
where ${N}_{src}$ is the source count per bin, $N_{bkg}$ is the background count per bin and $A_{src}/A_{bkg}$ is the ratio of the areas of the source and background regions. The significance of the overdensity is calculated per color bin and the maximally overdense bin is identified.
To accurately estimate the over-density peak, we use bin sizes of 0.05, 0.1 and 0.15 for the color distribution, with and without a half shift in the bin center. Once we determine the maximally overdense bin, our algorithm incorporates other neighbouring bins with excess galaxy counts to determine the peak of the overdensity. The criterion for including the neighbouring bins is set as $N_{src}/N_{bkg} > 2$. The bins that satisfy the aformentioned criteria are combined together to determine the mean color of the red sequence and the color error is given by the standard deviation. The significance of detection is  calculated using the total source and background counts for all excess bins. The color of the cluster is converted to redshift using the color-redshift relation found by fitting a broken power-law to the spectroscopic data for $55,000 $\ rich clusters from the GMBCG catalog \citep{hao10}, spanning a redshift range of $0.1 < z < 0.55$ (See Figure~\ref{image3}). The broken power laws for $g-r$ and $r-i$ display  a break-point at $z=0.35$. For photometric redshift estimation, $g-r$ colors have been used to identify clusters with $z<0.35$ due to the flatness of the relation at higher redshifts. While r$-$i colors shows a relatively steeper trend in both redshift intervals, we have predominantly used the $r-i$ colors for the redshift range $0.35<z<0.7$. Because of the flat relation of colors as a function of redshift for $i-z$ and $z-y$, we have used the redsequence in these color bands only for detection purposes and not for redshift detemination. The uncertainities in the color are converted to redshift using the propogation of errors and  combined in quadrature with the scatter in the color-redshift relation. The redshift estimates using the red sequence method are reported in Table~1.

\begin{figure}
    \includegraphics[width=0.5\textwidth]{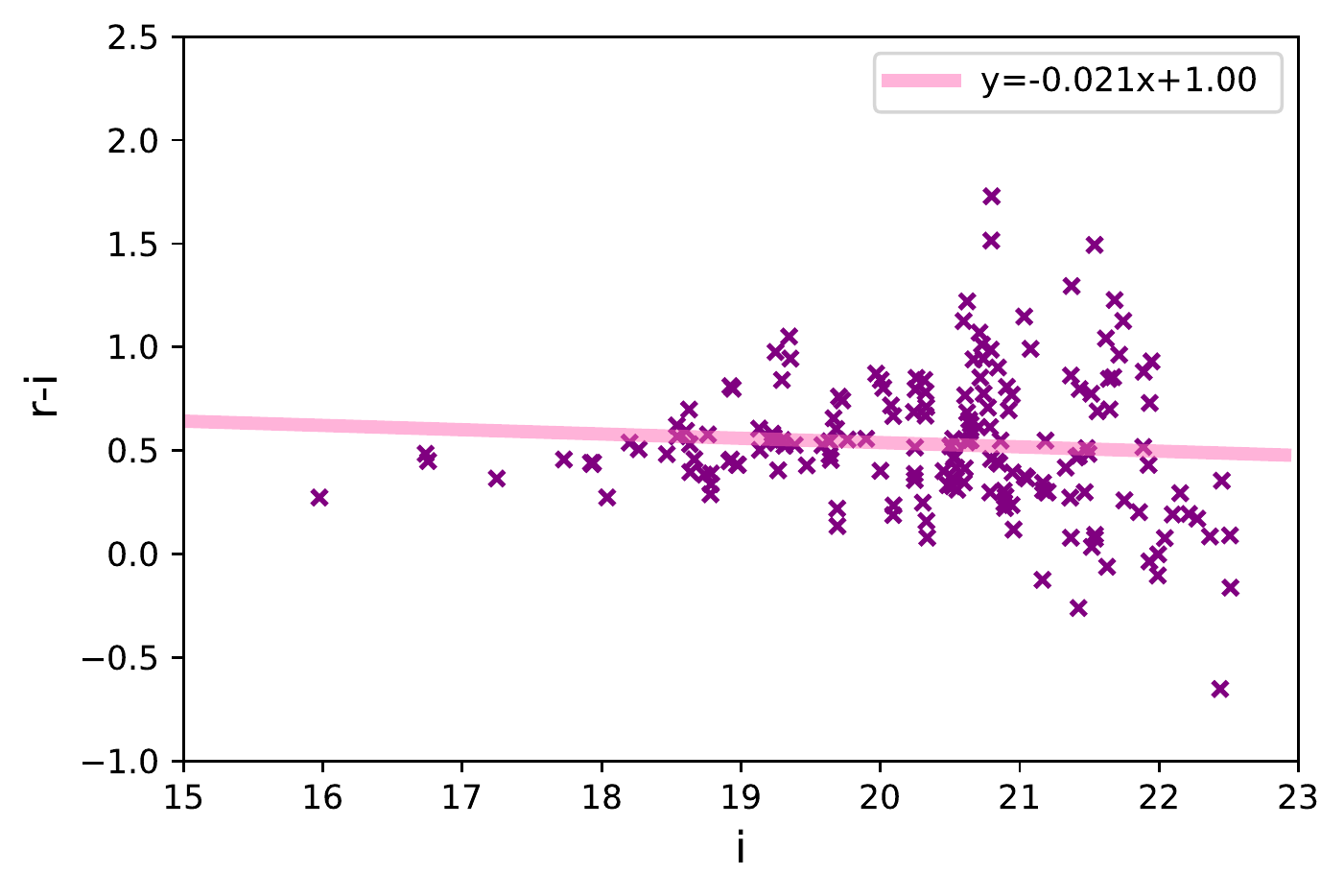}
    %{SWCL_J021007.7-270414_redseq.pdf}
    \includegraphics[width=0.5\textwidth]{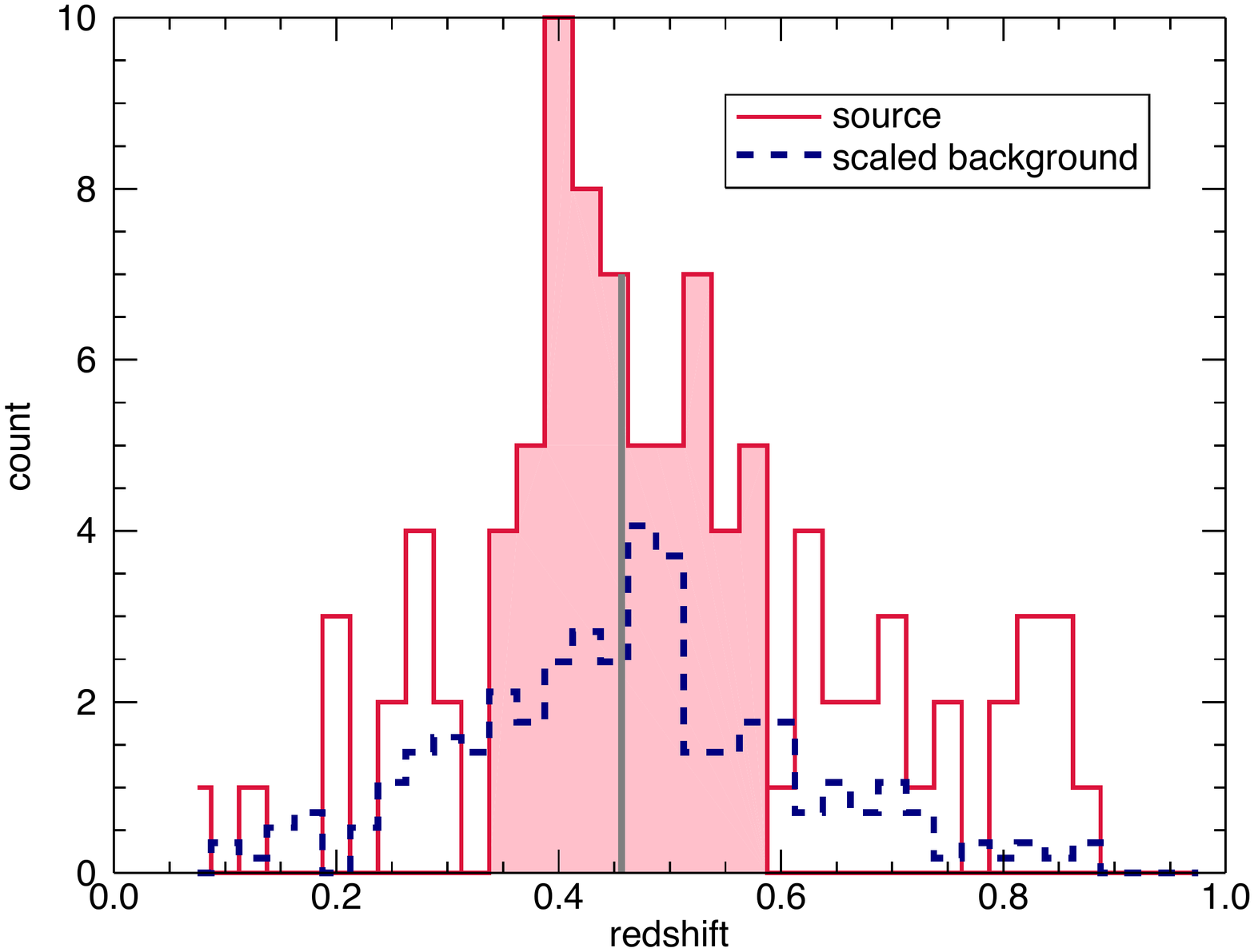}
    \includegraphics[width=0.5\textwidth]{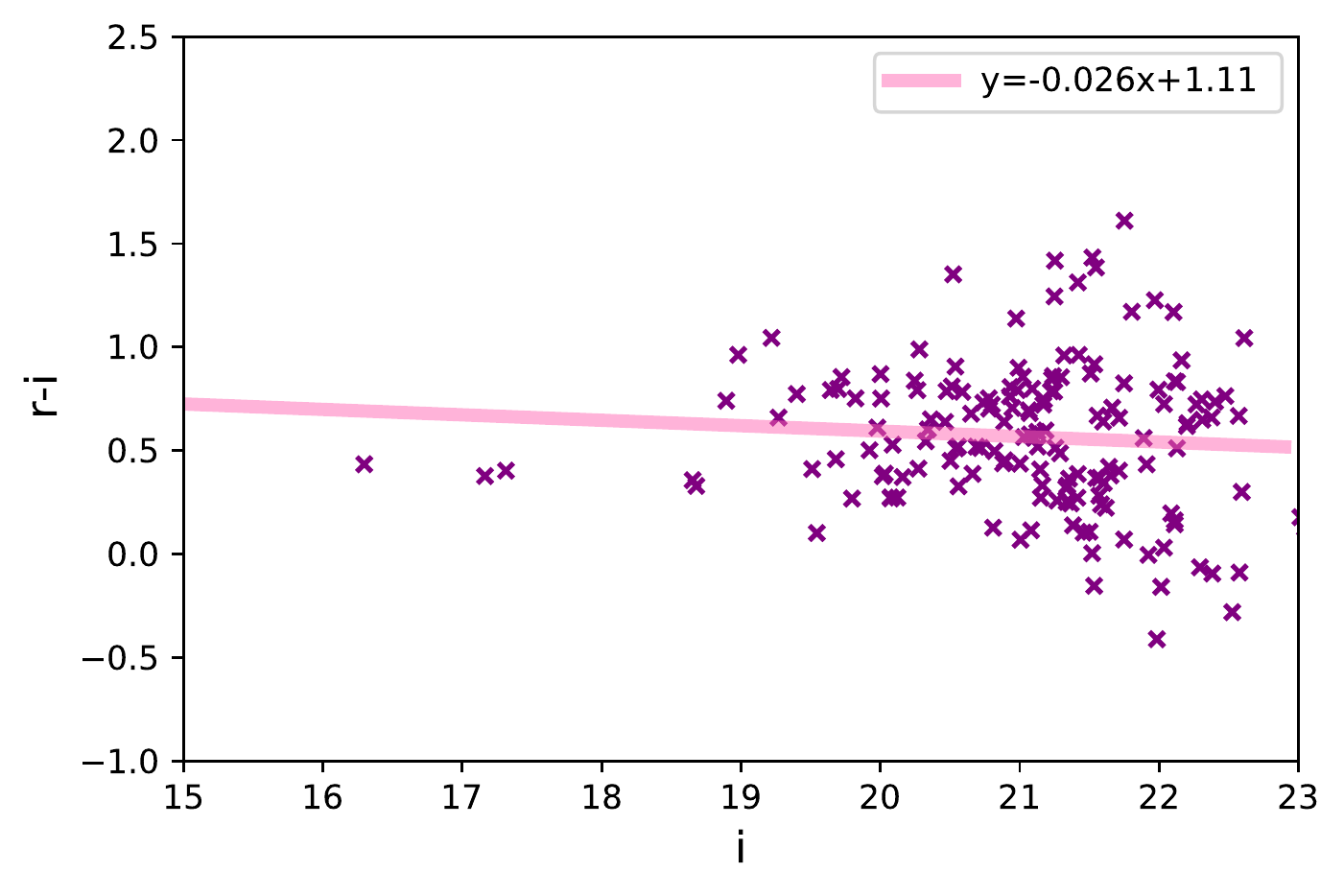}
    \includegraphics[width=0.5\textwidth]{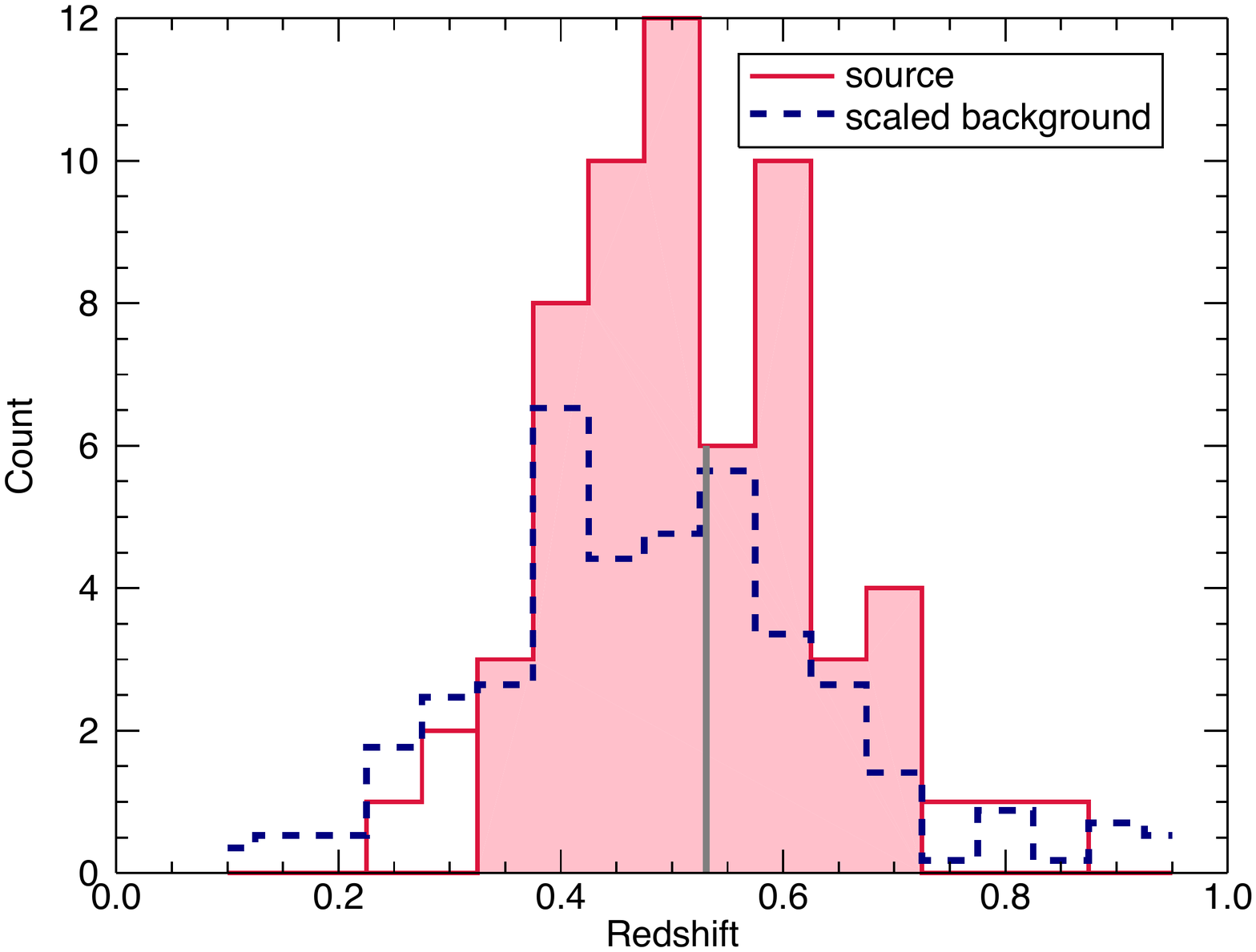}
    \caption{The color-magnitude diagram (Left) and the photometric redshift distribution (Right) for galaxies within 1.5' of the cluster X-ray centroid for Swift sources SWCL J$002729.2-232626$ (Top) and SWCL J$021007.7-270414$ (Bottom). The scaled background galaxy distribution is represented by the navy dashed line. SWCL J$021007.7-270414$ (Top) is optically verified with a detection significance of 5.84$\sigma$ (Top right) and redshift of $0.45$. SWCL J$002729.2-232626$ detected at a redshift of 0.54 and detection significance of 3.35$\sigma$ (Bottom right). The redshift bins with significant overdensities are shaded in red and the mean redshift is shown by a gray vertical line.
    \label{fig:image2}}
\end{figure}
\subsection{Redshift estimation using EAZY photo-zs}
Apart from looking for clustering in the color space, we have also run a similar analysis to locate overdensities in redshift space. This method requires photo-z estimation of the galaxies, which has been conducted using the photo-z estimator EAZY \citep{bram08}. EAZY employs a spectral energy distribution (SED) fitting technique to compute the photometric redshifts of galaxies using broadband photometry, and provides reasonably accurate photo-z estimates without the need for spectroscopy. The accuracy of the photo-z estimates depend on a number of factors, one of them being the availability of multiband imaging data in 5 or more filters, therefore the photo-z estimates were obtained only for the \pan\ data which provides imaging in g, r, i, z and y band. The photo-z redshift distribution of the source and the background galaxies yield the mean photometric redshift (See Figure~\ref{fig:image2}) of the cluster and the detection significance. For the redshift, we have used the same  algorithm as used in color space to determine the mean redshift and the significance of the detection. An average of the photo-z errors for the galaxies are combined in quadrature with the standard deviation of the mean redshift to find the uncertanities. Owing to the uncertainties in the redshift measurements, some clusters with an overdensity in color space may not present a counterpart detection in the redshift space, therefore we have reported the candidates that satisfy the detection criteria for either one of the cluster-finding methods. We require a $>3\sigma$ overdensity for a detection, however, if both give a $>3\sigma$ detection, the highest detection significance among the two is considered and the corresponding redshift estimates are used. In Table~\ref{table1}, we lists 55 Swift clusters that are confirmed with a detection above $3\sigma$. We have reported the redshift estimates using the red sequence and the photo-z method. In Figure~\ref{fig:image2}, we show the color magnitude diagram and galaxy redshift distribution for two detected SACS clusters. 

\begin{figure}
 \includegraphics[width=\textwidth]{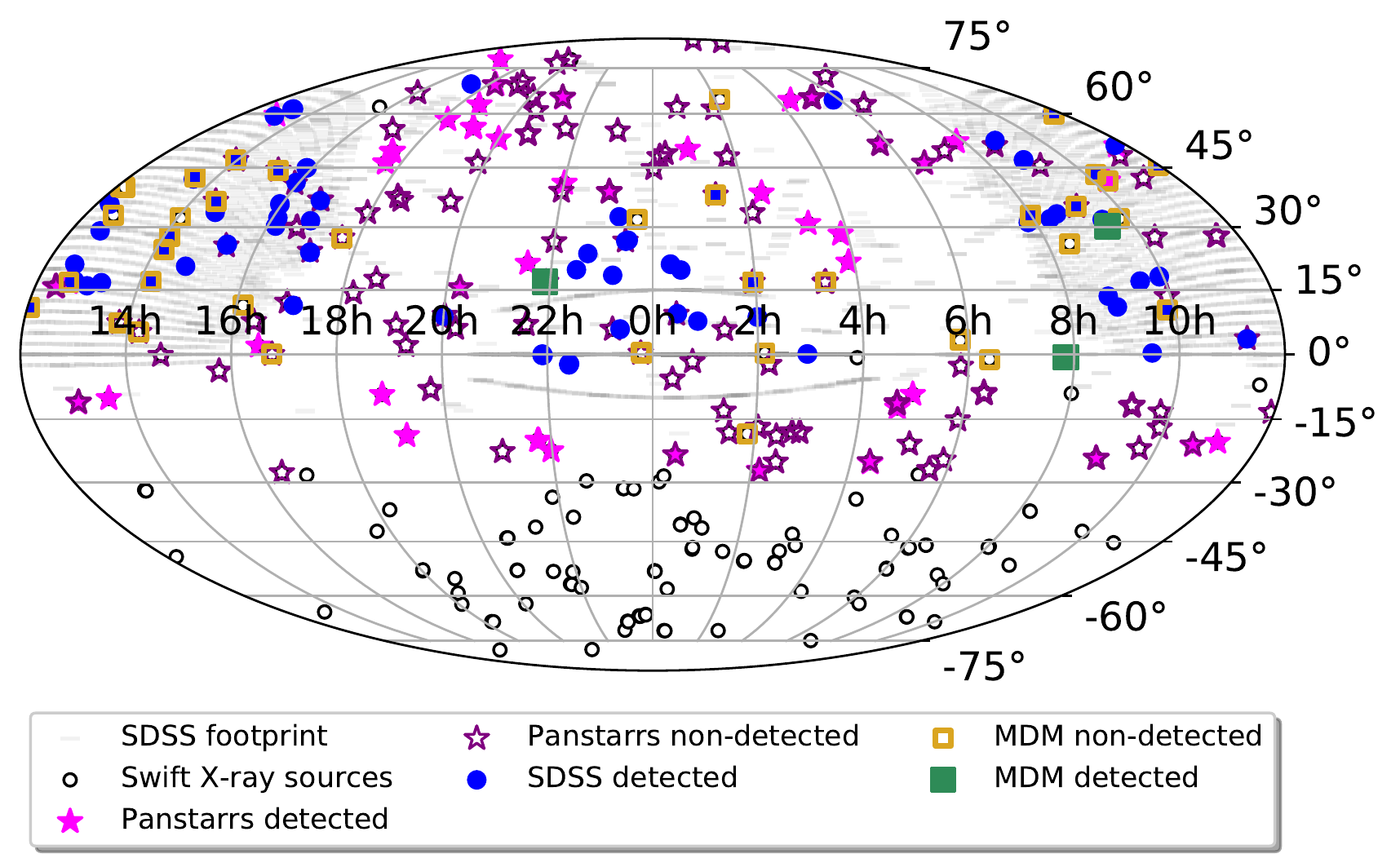}
    \caption{Equatorial coordinate distribution of the 442 SACS cluster candidates. The black open circles are those without follow-up data from this paper. The SDSS spectroscopic plates for the DR8 are shown in gray and the SDSS detections \citep{griff16} are displayed as blue filled circles. The Swift cluster candidates falling within the SDSS footprint and followed up with deeper optical observations with MDM are shown as squares, with the detected clusters as filled green and undetected clusters as open golden symbols. The Swift cluster candidates falling outside the SDSS footprint analyzed using Pan-STARRS data are marked by filled pink and open purple stars for detections and non-detections respectively.
    \label{image4}}
\end{figure}

\startlongtable
%\begin{rotatetable}
%\begin{deluxetable*}{ccCrlc}
\begin{deluxetable*}{ccccccccccccc} %{cDDDDDD}
\tabletypesize{ \scriptsize}%\footnotesize}
\tablecaption{Swift Cluster Survey: $ > 3\sigma$ optical confirmations\label{table1}}
\tablecolumns{13}
\tablenum{1}
\tablewidth{9pt}
%\label{tab:1}
%\tablewidth{50pt}
\tablehead{
\colhead{ Swift name} &
\colhead{RA} &
\colhead{Dec} &
\colhead{Detection} &
\colhead{z} &
\colhead{$\sigma_{z}$} &
\colhead{z} &
\colhead{$\sigma_{z}$} &
\colhead{$\lambda$} &
\colhead{$N_{o}$} &
\colhead{$\sigma_{\lambda}$} &
\colhead{Lx } &
\colhead{$\sigma_{L_X}$ }\\
\colhead{} &
\colhead{($deg$)} &
\colhead{($deg$)}&
\colhead{significance}&
\colhead{(Color)}&
\colhead{}&
\colhead{(EAZY)}&
\colhead{}&
\colhead{}&
\colhead{}&
\colhead{}&
\colhead{(ergs/s)} &
\colhead{(ergs/s)}
}
%\decimels
\startdata
%\tablenotemark{a} Out of the 52 Swift cluster candidates analyzed with the MDM data, 16 are detected at a significance of > 2 $\sigma$
SWCL J051046.0+644429  & 77.69 & 64.74 & 3.79 &     0.17 & 0.042 & 0.44 & 0.129 &    .... & 6.06 & .... &    4.30e+42 & 4.02e+41 \\ 
SWCL J104158.8$-$211124  & 160.49 & $-$21.19 & 3.68 &     0.18 & 0.173 & 0.45 & 0.100 &    14.71 & 14.12 & 3.757 &     3.23e+43 & 4.05e+42 \\ 
SWCL J183744.0+624135  & 279.43 & 62.69 & 4.74 &     0.20 & 0.037 & 0.44 & 0.112 &    9.58 & 14.41 & 3.796 &     7.25e+43 & 1.45e+43 \\ 
SWCL J054653.1+510908  & 86.72 & 51.15 & 4.76 &     0.20 & 0.173 & 0.42 & 0.096 &    12.83 & 13.47 & 3.670 &     5.21e+43 & 9.92e+42 \\ 
SWCL J133437.4$-$100927  & 203.66 & $-$10.16 & 3.27 &     .... &  .... & 0.79 & 0.486 &    117.97 & 2.12 & 1.455 &     1.02e+45 & 8.74e+43 \\ 
SWCL J062915.2+460619  & 97.31 & 46.11 & 5.60 &     0.17 & 0.036 & 0.40 & 0.097 &    15.88 & 25.29 & 5.029 &     2.29e+45 & 8.01e+43 \\ 
SWCL J181053.5+581524  & 272.72 & 58.26 & 5.52 &     0.36 & 0.036 & 0.39 & 0.079 &    7.31 & 12.82 & 3.581 &     2.23e+44 & 2.32e+43 \\ 
SWCL J155644.8+782352  & 239.19 & 78.40 & 3.56 &     .... &  .... & 0.40 & 0.121 &    .... & 6.12 & .... &     1.19e+43 & 2.30e+42 \\ 
SWCL J173719.2+461253  & 264.33 & 46.21 & 6.53 &     0.40 & 0.038 & 0.55 & 0.099 &    64.16 & 41.59 & 6.449 &     9.21e+43 & 6.45e+42 \\ 
SWCL J173721.7+461832  & 264.34 & 46.31 & 6.56 &     0.49 & 0.039 & 0.66 & 0.124 &    123.15 & 17.82 & 4.222 &     3.47e+44 & 1.49e+43 \\ 
SWCL J035130.3+281517  & 57.88 & 28.25 & 4.02 &     0.06 & 0.035 & 0.40 & 0.098 &    .... & 8.94 & .... &    5.28e+41 & 8.58e+40 \\ 
SWCL J002729.2$-$232626  & 6.87 & $-$23.44 & 3.97 &     0.44 & 0.035 & 0.52 & 0.108 &    5.32 & 8.00 & 2.828 &     1.31e+44 & 1.10e+43 \\ 
SWCL J005233.8+495407  & 13.14 & 49.90 & 3.51 &     0.48 & 0.035 & 0.40 & 0.104 &    .... & 2.53 & .... &    7.14e+44 & 3.47e+43 \\ 
SWCL J021007.7$-$270414  & 32.53 & $-$27.07 & 7.13 &     0.42 & 0.037 & 0.52 & 0.106 &    76.62 & 44.35 & 6.660 &     2.96e+43 & 2.33e+42 \\ 
SWCL J022409.4+382635 & 36.04 & 38.44 & 4.62 &     .... &  .... & 0.74 & 0.222 &   .... & .... & .... &   .... & .... \\ 
SWCL J031430.6+305035  & 48.63 & 30.84 & 3.07 &     0.62 & 0.038 & ....  & .... &    96.96 & 3.18 & 1.782 &    6.59e+43 & 9.03e+42 \\ 
SWCL J035312.2+213345  & 58.30 & 21.56 & 9.17 &     0.17 & 0.040 & 0.48 & 0.103 &    55.38 & 73.41 & 8.568 &     1.14e+44 & 9.08e+42 \\ 
SWCL J042338.6$-$251617  & 65.91 & $-$25.27 & 3.07 &     0.26 & 0.040 & 0.53 & 0.144 &    .... & 6.53 & .... &    3.48e+42 & 6.47e+41 \\ 
SWCL J042422.3+640633  & 66.09 & 64.11 & 3.30 &     0.55 & 0.041 & 0.53 & 0.073 &    170.64 & 3.94 & 1.985 &     5.66e+43 & 1.16e+43 \\ 
SWCL J044123.7$-$111550  & 70.35 & $-$11.26 & 4.05 &     0.55 & 0.036 & 0.51 & 0.100 &    16.35 & 12.59 & 3.548 &     3.17e+43 & 5.16e+42 \\ 
SWCL J044144.6$-$111534 & 70.44 & $-$11.26 & 3.05 &     .... & ....  & .... & .... &         .... & .... & .... &   .... & .... \\ 
SWCL J044237.2$-$122251  & 70.66 & $-$12.38 & 3.74 &     0.11 & 0.173 & 0.44 & 0.106 &    .... & 12.06 & .... &    1.39e+42 & 2.90e+41 \\ 
SWCL J045832.6$-$091111  & 74.64 & $-$9.19 & 5.83 &     0.49 & 0.037 & 0.28 & 0.132 &    16.81 & 6.53 & 2.555 &     2.75e+44 & 1.80e+43 \\ 
SWCL J074755.5+515852  & 116.98 & 51.98 & 5.01 &     0.51 & 0.036 & 0.68 & 0.088 &    25.64 & 5.41 & 2.326 &     1.50e+44 & 2.59e+43 \\ 
SWCL J085416.4$-$240703  & 133.57 & $-$24.12 & 3.08 &     0.36 & 0.037 & 0.42 & 0.116 &    7.51 & 10.71 & 3.272 &     7.65e+44 & 1.42e+43 \\ 
SWCL J102036.8+413227  & 155.15 & 41.54 & 3.05 &     0.36 & 0.036 & 0.42 & 0.229 &    .... & 5.18 & .... &    1.91e+43 & 3.42e+42 \\ 
SWCL J110932.9$-$202209  & 167.39 & $-$20.37 & 3.34 &     0.05 & 0.036 & 0.73 & 0.368 &    10.64 & 6.35 & 2.521 &     1.03e+44 & 1.48e+43 \\ 
SWCL J114332.8+504856  & 175.89 & 50.82 & 5.43 &     0.35 & 0.036 & 0.42 & 0.101 &    5.55 & 8.94 & 2.990 &     1.43e+43 & 2.66e+42 \\ 
SWCL J122327.6+153927  & 185.86 & 15.66 & 4.75 &     0.29 & 0.173 & 0.45 & 0.123 &    5.27 & 7.53 & 2.744 &     3.26e+43 & 5.76e+42 \\ 
SWCL J123717.7+164353  & 189.32 & 16.73 & 3.07 &     0.13 & 0.173 & ....  & .... &    .... & 4.06 & .... &    6.40e+42 & 2.95e+41 \\ 
SWCL J125814.0$-$111333  & 194.56 & $-$11.23 & 3.54 &     0.40 & 0.036 & 0.39 & 0.077 &    4.85 & 8.53 & 2.921 &     1.62e+43 & 3.09e+42 \\ 
SWCL J130345.6+593437  & 195.94 & 59.58 & 4.44 &     0.23 & 0.039 & 0.42 & 0.092 &    5.04 & 7.82 & 2.797 &     1.49e+43 & 2.43e+42 \\ 
SWCL J163054.8+015924 & 247.73 & 1.99 & 7.21 &     .... & ....  & .... & .... &         .... & .... & .... &   .... & .... \\ 
SWCL J173302.3+490920  & 263.26 & 49.16 & 4.58 &     0.43 & 0.036 & 0.45 & 0.120 &    4.12 & 6.35 & 2.521 &     1.63e+43 & 2.95e+42 \\ 
SWCL J173316.3+492211  & 263.32 & 49.37 & 3.08 &     0.47 & 0.038 & ....  & .... &    .... & .... & .... &    3.01e+43 & 4.53e+42 \\ 
SWCL J181628.8+691131  & 274.12 & 69.19 & 6.63 &     0.14 & 0.038 & 0.45 & 0.116 &    13.79 & 19.76 & 4.446 &     2.77e+44 & 1.54e+43 \\ 
SWCL J184929.4$-$091328  & 282.37 & $-$9.22 & 4.68 &     0.23 & 0.037 & 0.40 & 0.131 &    .... & 22.24 & .... &    1.07e+44 & 1.71e+43 \\ 
SWCL J190614.5+555534  & 286.56 & 55.93 & 8.31 &     0.27 & 0.043 & 0.49 & 0.097 &    69.37 & 64.59 & 8.037 &     5.36e+44 & 2.78e+43 \\ 
SWCL J190620.9+555237  & 286.59 & 55.88 & 4.20 &     0.36 & 0.035 & 0.48 & 0.101 &    24.67 & 14.71 & 3.835 &     1.89e+44 & 2.04e+43 \\ 
SWCL J190633.9+560146  & 286.64 & 56.03 & 3.92 &     0.47 & 0.036 & ....  & .... &    .... & 4.35 & .... &    2.61e+43 & 5.50e+42 \\ 
SWCL J191020.9$-$184932  & 287.59 & $-$18.83 & 4.02 &     0.08 & 0.037 & 0.42 & 0.150 &    .... & 13.76 & .... &    3.70e+42 & 6.51e+41 \\ 
SWCL J200005.7+524438 & 300.02 & 52.74 & 3.32 &     .... & ....  & .... & .... &         .... & .... & .... &   .... & .... \\ 
SWCL J201549.0+153231  & 303.95 & 15.54 & 3.26 &     0.04 & 0.036 & 0.40 & 0.165 &    .... & 23.47 & .... &    1.94e+41 & 2.47e+40 \\ 
SWCL J210442.9+644555  & 316.18 & 64.77 & 3.91 &     0.52 & 0.036 & 0.52 & 0.092 &    68.87 & 6.18 & 2.485 &     3.02e+43 & 3.94e+42 \\ 
SWCL J213130.8+211616  & 322.88 & 21.27 & 3.05 &     0.59 & 0.037 & ....  & .... &    .... & .... & .... &    1.78e+44 & 1.78e+43 \\ 
SWCL J214405.7$-$195813  & 326.02 & $-$19.97 & 7.77 &     0.49 & 0.038 & 0.60 & 0.135 &    91.90 & 33.00 & 5.745 &     4.55e+43 & 6.42e+42 \\ 
SWCL J214409.9$-$195600  & 326.04 & $-$19.93 & 3.34 &     0.72 & 0.036 & 0.82 & 0.337 &    .... & 9.12 & .... &    3.84e+45 & 8.45e+43 \\ 
SWCL J214515.6$-$195944  & 326.32 & $-$20.00 & 3.91 &     0.04 & 0.036 & 0.50 & 0.144 &    .... & 8.00 & .... &    6.85e+42 & 1.56e+41 \\ 
SWCL J215507.7+164725  & 328.78 & 16.79 & 3.50 &     0.11 & 0.173 & 0.72 & 0.165 &    .... & 10.88 & .... &    5.10e+43 & 4.42e+42 \\ 
SWCL J215831.9$-$222439  & 329.63 & $-$22.41 & 6.86 &     0.38 & 0.037 & 0.45 & 0.106 &    15.84 & 22.82 & 4.777 &     1.50e+43 & 2.94e+42 \\ 
SWCL J220026.5+405625  & 330.11 & 40.94 & 3.18 &     0.25 & 0.173 & ....  & .... &    .... & 4.94 & .... &    3.22e+43 & 1.73e+42 \\ 
SWCL J230207.3+384751  & 345.53 & 38.80 & 4.94 &     0.59 & 0.037 & 0.37 & 0.129 &    6.48 & 12.47 & 3.531 &     5.81e+43 & 5.43e+42 \\ 
SWCL J092642.2+300835  & 141.68 & 30.14 & 3.01 &     0.90 & 0.045 & ....  & .... &    128.30 & 9.40 & 3.066 &    8.39e+43 & 1.35e+43 \\ 
SWCL J075036.6$-$003838  & 117.65 & $-$0.64 & 3.43 &     0.09 & 0.036 & ....  & .... &    .... & 10.16 & .... &    7.38e+43 & 5.21e+42 \\ 
SWCL J215357.0+165313  & 328.49 & 16.89 & 4.28 &     0.43 & 0.036 & ....  & .... &    .... & 12.76 & .... &    2.74e+44 & 1.22e+43 \\ 
\enddata
%\tablecomments{Out  of 51 Swift clusters analyzed with the MDM optical data, 22 were detected with a significance of $> 2 \sigma$}
%\begin{tablenotes}
%      \small
%      \item Out of 52 Swift clusters  analyzed with the MDM optical data, 16 were detected with a significance of $> 2 \sigma$. 
%    \end{tablenotes}
\end{deluxetable*}
%\end{rotatetable}

\begin{figure}
    \includegraphics[width=\textwidth]{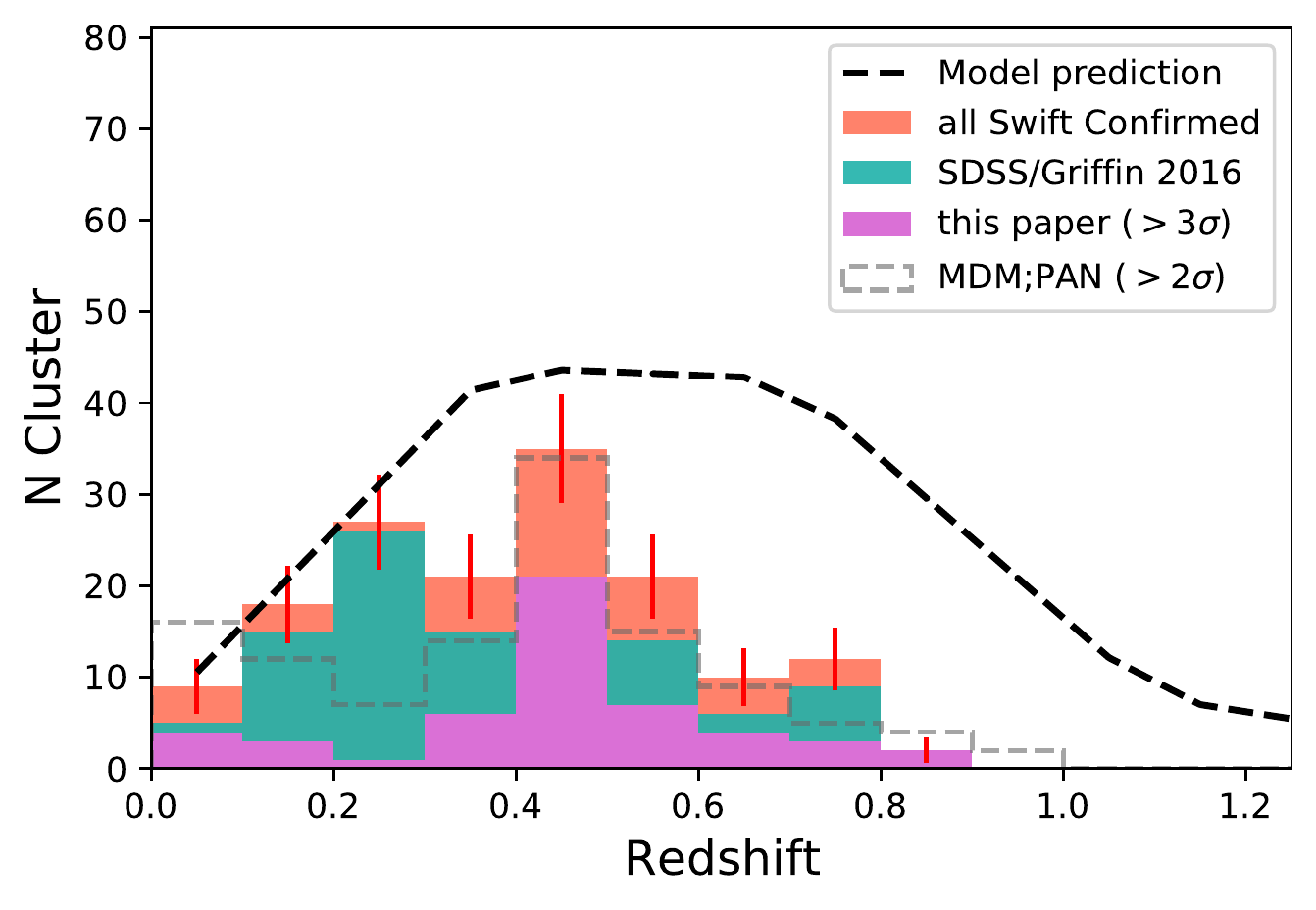}
\caption{Redshift distribution of the optically confirmed SACS clusters detected at a significance $> 3\sigma$. The green histogram is the z distribution of the SDSS confirmed clusters from \citep{griff16}. The pink histogram is the redshift distribution of the clusters optically detected at $> 3\sigma$ in this paper using MDM and Pan-STARRS, and the peach histogram is the distribution of all optically confirmed clusters in the survey to date. The gray dashed histogram shows the distribution of all  $>2\sigma$ SACS candidates in MDM and Pan-STARRS. The black dashed line shows the predicted distribution for the Swift AGN and cluster survey using the model of \citet{tink08}.
\label{fig:image5}}
\end{figure}

\begin{figure}
    \includegraphics[width=0.5\textwidth]{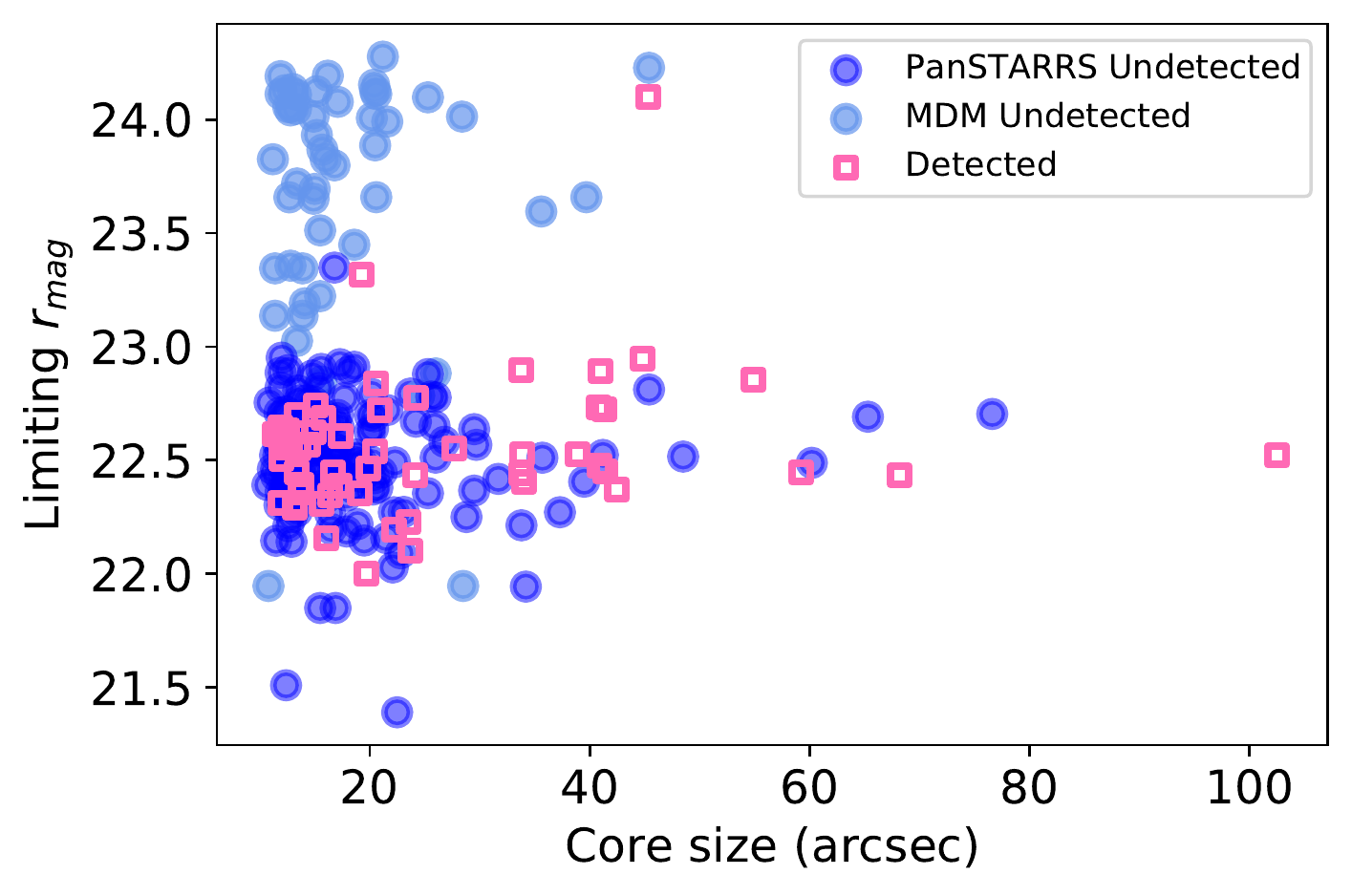}
    \includegraphics[width=0.5\textwidth]{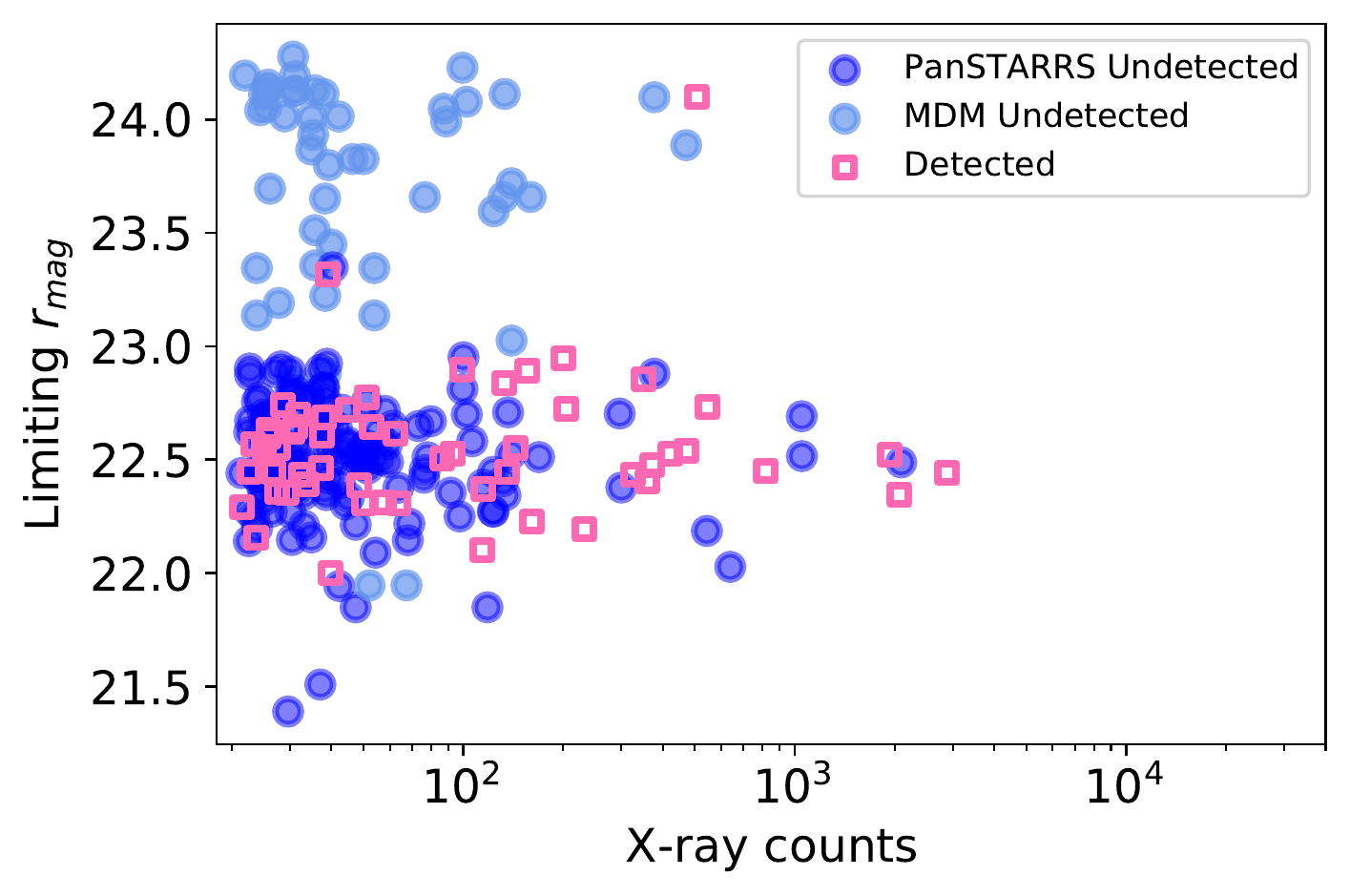}
    \includegraphics[width=0.5\textwidth]{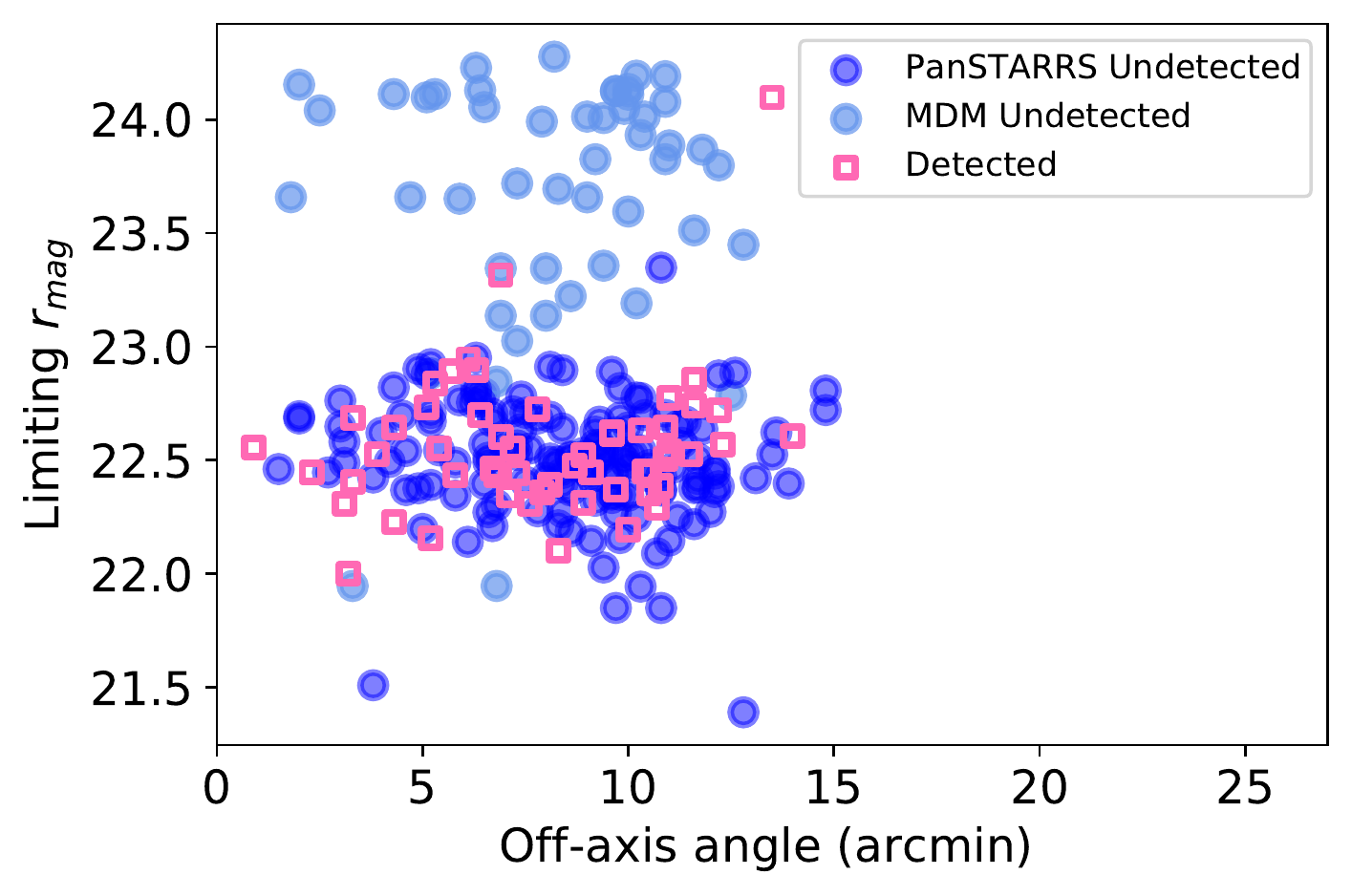}
    \includegraphics[width=0.5\textwidth]{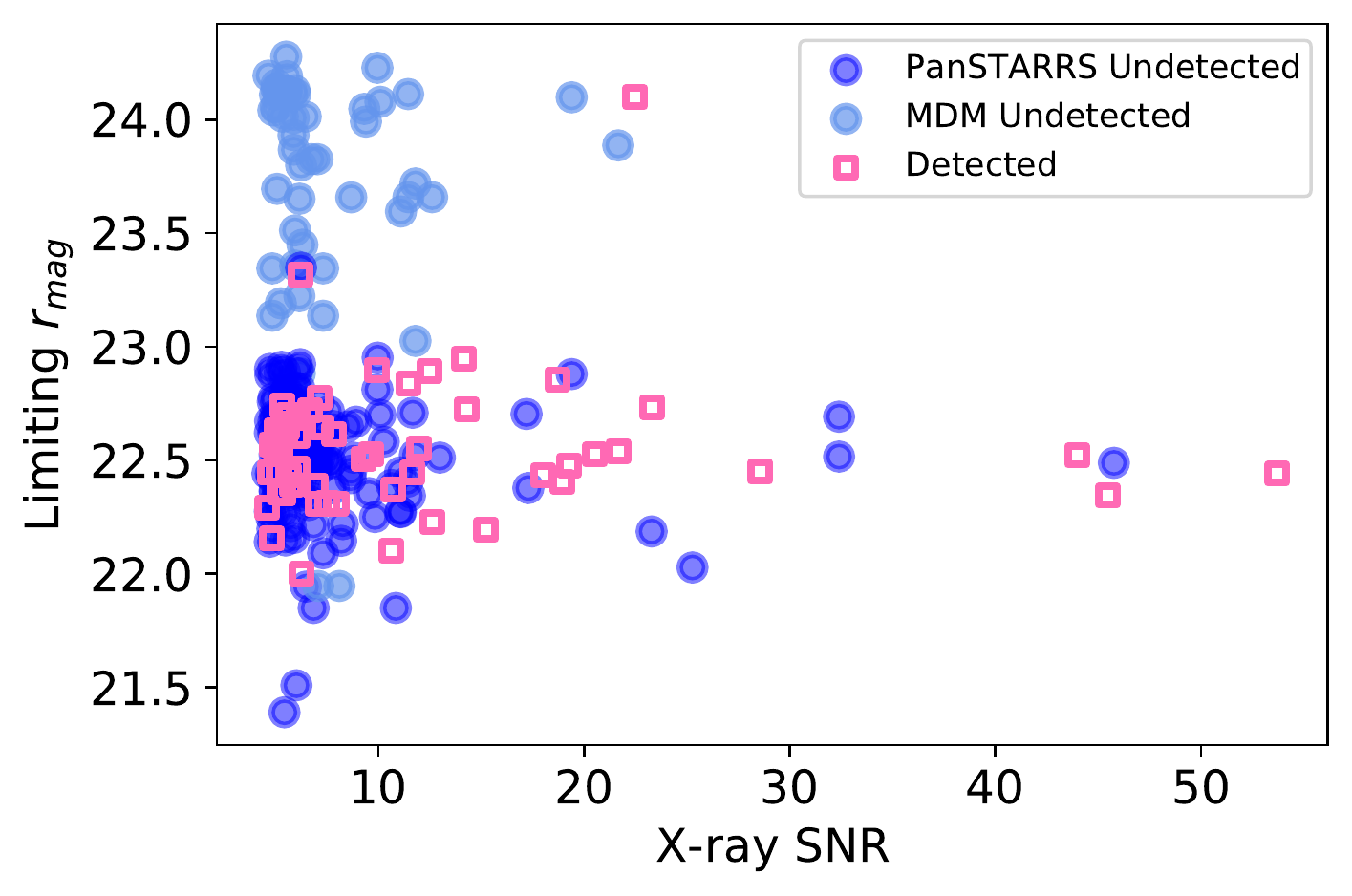}
    \caption{Distribution of the candidates examined here in limiting r-band magnitude on X-ray core-size (top left), X-ray counts (top right), off-axis angle (lower left) and X-ray SNR (lower right). The candidates are coded as shown in the panels.
    \label{image6}}
\end{figure}

\subsection{X-ray luminosity and Optical richness}

We also report the X-ray luminosity and the optical richness for all the detected clusters in the Table~1. A detailed study of the X-ray and optical properties and their correlations for the SACS clusters will be studied in the subsequent paper. For the calculation of X-ray bolometric luminosities, we have utilized an X-ray spectral fitting program XSPEC \citep{arn96}. We used the flux estimates for the SACS clusters from \citet{dai15} and converted the flux to luminosity by assuming a multiplicative component model, $wabs*apec$. The plasma temperature was fixed at 5 keV and the abundance is assumed to be 0.3 Solar. The Galactic column density was fixed for each cluster position using the n$_{H}$ command in XSPEC. We use the photometric redshifts from our present analysis. The uncertainties in L$_{X}$ only include the uncertainties in the X-ray photon counts and not the model parameters. Since X-ray luminosity serves as a mass proxy, we have used the M$_{200}$--L${_X}$ relation \citep{rei02} to estimate the mass within the radius of R$_{200}$ at which the density of the cluster is 200 times the critical density of the Universe.
%These estimates are further used for the calculation of the optical richness, discussed below. 

The optical richness of the cluster, $\lambda$, is a measure of the number of galaxies in the system. To estimate $\lambda$, we first measure the observed galaxy counts, $N_{o}$, which is the number of galaxies above the estimated background that fall within the one standard deviation of the mean redshift of the cluster. The  optical richness, $\lambda$, is the number of galaxies with luminosities larger than $L_{*}$ or magnitude brighter than $M_{*}$. We assume a Schechter luminosity function, 
\begin{equation}
     \lambda=\int^{\infty}_{L_{*}} \ \phi_{*} \ {\Bigg( \frac{L}{L_*} \Bigg)}^{\alpha} \ e^{-\frac{L}{L_*}} \ dL  
\end{equation}
where,
\begin{equation}
\frac{L}{L_*}= 10^{0.4(M_* \ -\ M)}
\end{equation}
We have considered a magnitude break at $M_{*}(0)=-21.34$ mag and slope $\alpha = -1.07$, adopted from the results of \citep{bel03} for SDSS-r band. We assume that the break luminosity evolves as
\begin{equation}
M_{*}(z) = M_{*}(0) − \beta z
\end{equation}
Where $\beta = 1.2$ \citep{dai09}. In order to normalize the Schechter luminosity function, we determine the absolute magnitude limits for each cluster using the apparent r-band limiting magnitudes for each field. For the calculations of the absolute magnitude limits, we correct for galactic dust extinction for each cluster centroid position using the NED online calculator for Galactic Reddening and Extinction. We also apply the K-corrections determined using the low resolution spectral templates for elliptical galaxies from \citet{ass10}. The normalization constant for the Schechter luminosity function, $\phi_{*}$ , is calculated using the absolute magnitude limit and the background subtracted source counts, $N_{o}$. Considering $N_{o}$ is found using an apparent radius of $\ang{;3;}$, we make a correction for $\lambda$ assuming an aperture radius of 1.0 Mpc. We adopt an NFW density profile \citep{nav95} to calculate the correction factor, 
which is the ratio of the density within a projected radius of 1 Mpc $\rho(r < R1$) to the density within the observed radius $\rho( r < R_{obs})$. Here, $R_{obs}$ is the angular diameter distance in Mpc corresponding to the $\ang{;3;}$ of the source region. We estimated the NFW scale radius r$_s$ using the L$_X$ -- M$_{200}$ and M$_{200}$-- c$_{200}$ relations from \citet{rei02} and \citet{ett10}. The uncertainty in $\lambda$ heavily depends on the uncertainties in the background subtracted number counts, which is a given as the Poisson error $\sqrt N_{o}$. For several cases with redshifts $<0.1$, the optical richness estimates are severely underestimated due to the missing galaxies because of the poor image quality of the observations. On the other hand, for higher redshifts $z>0.5$, we see that optical richness is grossly overestimated, which has been corrected by setting a magnitude cut-off at $M_{*}-1.5 \ mag$. We have found that the changes in the magnitude limits have a systematic effect on the richness estimates as we traverse from the fainter to the brighter end of the luminosity function. The observed galaxy counts and the optical richness estimates are presented in Table 1, and we have not reported the values for under estimated clusters with $\lambda <4$.
\newpage

\section{Results and Discussion} \label{sec:discuss}
%In this paper, we discuss the red sequence detection technique that is motivated by the tight color magnitude relation of the red elliptical galaxies in galaxy clusters. We discuss the origin and evolution of the CMR, and also describe the various observational programmes that have extended the red sequence studies to higher redshifts. The Swift X ray and cluster survey is also among these observational programmes; It is a serendipitous survey with an area and depth favourable for identifying clusters upto z$\sim$1. It is expected that the final survey of the Swift X ray telescope will provide a largest X ray selected cluster catalog with $\sim$1000 clusters. In our paper, we use the Swift catalog from Data Release 1 consisting of 442 cluster candidates that need to be optically verified. Among these 442 cluster candidates, 203 cluster candidates lie in the domain of the SDSS and have been studied by Griffin et al. 2016. They have optically confirmed 104 of 203 cluster candidates and have estimated the Swift catalog to be 80\% complete for low redshift clusters extending up to z=0.4.  As for the remaining unconfirmed clusters, they expect them to be high redshift clusters that are beyond the scope of the SDSS, and proposed deeper optical observations. 

This is the second paper in a series searching for optical counterparts and estimation of redshifts for the SACS X-ray survey. \citet{griff16} identified 104 of the SACS clusters using SDSS DR8. Here we identify another 55 clusters North of declination $\delta=-30^{\circ}$ using MDM and PAN-STARRS data. The next paper will cover $\delta <-30^{\circ}$ using CTIO and DES data. All the confirmations to date are illustrated in Figure~\ref{image4}. The confirmed clusters from this work with $>3\sigma$ overdensities extend up to $z\sim1$ with the majority of detections ranging within the redshift of $0.3-0.6$. Figure~\ref{fig:image5} shows the redshift distribution of all the optically confirmed SACS clusters detected at a $3\sigma$ significance threshold. We also show the theoretical expectations derived using  \citet{tink08} model for the mass function of dark matter halos and their redshift evolution. The  model assumes flat $\Lambda CDM$ cosmology with halo masses in the range $10^{14} h^{-1} M_\odot \leq M \leq 10^{15} h^{-1} M_\odot$. For this work, the model predictions are calculated assuming a slight change in cosmology with parameters: $\Omega_{m} = 0.25$, $\sigma_{8} = 0.9$,
$h = 0.72$, and $\Delta = 2000$ and masses ranging from $10^{14} h^{-1} M_\odot$ to $10^{15} h^{-1} M_\odot$. We have also taken into consideration the flux limit and the area of the Swift survey. This model provides a reasonable estimate for the expected distribution and allows us to test the completeness of the catalog. In Figure~\ref{fig:image5}, we compare the observed distribution with the model predictions. 

We find that the $>3\sigma$ overdensity sample is consistent with the theoretical redshift distribution and the SACS survey is complete up to $z\sim0.3$ and is nearly 80\% complete upto $z\sim 0.5$.  The number of detections show  a slight increase up to $z \sim 0.4 $, which is followed by a significant jump at $z \sim 0.45 $, and then a slow decline to as far as $z \sim 0.9 $. The $>2\sigma$ overdensity sample for this work is sizable and the redshift distribution of these cluster candidates is consistent with the model prediction (See Figure~\ref{fig:image5}). However, to test the robustness of our method and calibrate the detection significance, we ran our analysis on a sample of 100 random locations in the SDSS footprint. We ensured that these random locations were far removed from any known clusters within SDSS by comparing against the GMBCG -- DR7 \citep{hao10}, SACS -- DR8 \citep{griff16}, redMaPPer -- DR8 \citep{ryk14} catalogs. We found that a fair fraction of random sources displayed a $>2 \sigma$ significance and so for the sake of robustness we have used a $3\sigma$ detection threshold. 
Comparing the redshift distribution from this paper with the earlier Swift paper \citep{griff16}, we find that SDSS distribution peaks at $z \sim 0.3$ while the MDM/Pan-STARRS distribution shows a peak at $z \sim 0.45$, which has enhanced the overall number count in the redshift $0.4-0.6$ range. The SDSS distribution showed a redshift tail $0.6<z<0.8$, which is also observed in the MDM/PAN-STARRS distribution and extends upto $z \sim 0.9$. Despite of the deeper observations with MDM, we are unable to detect a higher number of clusters within the intermediate redshift range ($0.6 <z  <0.8$) which implies that most of the undetected clusters from \citet{griff16} are possibly higher redshift clusters. 
Unfortunately, neither PS1 nor MDM data were markedly deeper than SDSS, so only moderate progress was made towards completeness at intermediate redshifts. The  $z \gtrsim0.8$ clusters which should be 30\% of the sample requires near-IR follow-up observations, which will be published in a forthcoming paper.

We can further investigate the non-detections by comparing the properties of the optically detected and non-detected candidates to the optical limiting magnitudes and various X-ray properties, as shown in Figure~\ref{image6}, for the $>3 \sigma$ overdensity threshold sample. 
%The detectability of the X-ray clusters in the optical depends on 
The quality of the optical observations can be approximately by the average limiting magnitudes for the fields in the $r$ band.  For the X-ray properties, we examine the X-ray photon counts, S/N, emission core size, and off-axis angle in the \swift\ images. 
We used the Kolmogorov-Smirnov (K-S) test to statistically check whether the detected and undetected distributions are different. In Figure~6, the distribution of the limiting $r$ band magnitudes for the MDM clusters are distinguishably clustered around higher limiting magnitudes (in the upper left quadrant), therefore indicating that the non-detections produced from the deeper MDM observations are most likely high redshift clusters or false positives in the X-ray detection methods. However, because there are only three detections in the MDM sample, the K-S tests were applied only to the \pan\ targets.  The K-S test results show that the optically detected and non-detected targets are distinguished by X-ray photon counts and S/N with K-S null probability of $0.0043$ for both cases.  
This is expected since the high X-ray count or S/N clusters are more likely to be luminous and low redshift clusters. Even though we see that the clusters with large X-ray counts and high S/N are being found by the optical detection method, we do see some exceptions that could be higher redshift candidates or probably X-ray false positives.
For the core distribution, there is a definite suggestion that two samples are different (P=0.0405), which indicates that X-ray cluster candidates with larger core sizes are less likely to be false positives. 
The off-axis angle and limiting magnitudes distributions show no clear distinction between the two populations with null-probability of 0.581 and 0.954, respectively.
This confirms that the PSF of \swift\ is approximately uniform with respect to the off-axis angles.
While the limiting optical magnitude is important for optically confirming X-ray clusters, if the sample contains a large fraction of high redshift clusters where the red sequence moved to the NIR band, there will always be a large fraction of non-detections in the optical bands regardless of the limiting magnitudes, which is consistent with this result.  Therefore, we expect the unconfirmed cluster candidates left after the current papers to be higher redshift clusters that require follow-up observations in the near infrared or low luminosity intermediate redshift clusters that require significantly deeper optical observations.

We have also estimated the cluster observables like the optical richness and the X-ray luminosity for each confirmed clusters. Although this paper only reports the estimated values and not presents the scaling relationship between the observable properties, it can be generally stated that an increasing trend is observed in the richness of the clusters with the increase in the X-ray luminosity. We find that these clusters are predominantly located on the lower end of the richness relation with $\lambda < 25$ or they are rich clusters with $\lambda > 60$. For some cases however, the optical richness has been severely underestimated because of the missing galaxy counts at lower redshifts, which is possibly due to the poor image quality/seeing of the observations. Some higher redshift cases show an inflated estimate for the richness which has been corrected by imposing a luminosity cut. The high redshift clusters are more prone to projection effects and the net number counts can majorly impact the estimates for the richness. It is also important to consider that the photometric redshift estimates are subject to systematics and could be another factor leading to the underestimation/overestimation of the richness. A detailed analysis of the X-ray and optical observables and the scaling relations for all the confirmed SACS clusters, including those in the southern hemisphere south of declination of -30 degrees, will be studied in the next paper.

\begin{acknowledgments}
We are grateful to the anonymous referee for the helpful comments and recommendations.  We acknowledge the financial support from the NSF grant AST-1413056 and NASA ADAP program NNX17AF26G.
\end{acknowledgments}

\end{document}